\documentclass[journal]{IEEEtran}

\usepackage{graphicx}
\usepackage{amssymb}
\usepackage{amsthm}
\usepackage{stmaryrd}

\usepackage[usenames,dvipsnames,svgnames,table]{xcolor}

\usepackage{cite}
\usepackage{balance}
\usepackage{hyperref}
\hypersetup
{
    colorlinks=false,
    pdfborder={0 0 0},
}

\newcommand{\x}[1]{{x_{\textsc{#1}}}}
\newcommand{\BraceTwo}[1]{{\{\!\!\{\!#1\!\}\!\!\}}}

\usepackage{multirow}
\usepackage[cmex10]{amsmath}
\begin{document}

\title{Fast Encoding and Decoding of Flexible-Rate and Flexible-Length Polar Codes}%
\author{Muhammad~Hanif and Masoud Ardakani,~\IEEEmembership{Senior Member,~IEEE}%
\thanks{M. Hanif and M. Ardakani are with the Department of Electrical and Computer Engineering, University of Alberta, AB, Canada (email: mhanif@uvic.ca, ardakani@ualberta.ca).}
}%
\maketitle
\begin{abstract}
This work is on fast encoding and decoding of polar codes. We propose and detail 8-bit and 16-bit parallel decoders that can be used to reduce the decoding latency of the successive-cancellation decoder. These decoders are universal and can decode flexible-rate and flexible-length polar codes. We also present fast encoders that can be used to increase the throughput of serially-implemented polar encoders.
\end{abstract}
\section{Introduction}
\newtheorem{remark}{Remark}
\newtheorem{lemma}{Lemma}
\newtheorem{theorem}{Theorem}
\IEEEPARstart{P}{olar} codes are capacity-achieving block codes that are recently introduced by Ar{\i}kan \cite{PolarCodesSeminalWork}. Due to their provable capacity-achieving performance with low-complexity encoding and decoding, they have gained significant interest \cite{FastPolarDecoders,SarkisSystematicPolarCodes}. In particular, polar codes can be encoded and decoded in a recursive fashion, which results in encoding and decoding complexities of $\mathcal{O}(N \log_2{N})$, where $N$ is the code length \cite{PolarCodesSeminalWork}.

One of the main challenges associated with polar codes is their high decoding latency and low throughput \cite{UnrolledPolarDec,FastPolarDecoders}. Since the serial nature of decoding bottlenecks fast implementation of polar coding, researchers have introduced novel ways to reduce the decoding time \cite{SarkisSystematicPolarCodes,SimplifiedSCD,FastPolarDecoders,PolarCodesMagazine,UnrolledPolarDec}. For example, \cite{SimplifiedSCD} introduces the notion of rate-zero and rate-one nodes to reduce the decoding depth of the successive cancellation (SC) decoder. The resulting simplified SC (SSC) decoder reduces the decoding latency up to 20 times \cite{PolarCodesMagazine}. The decoding latency can further be improved by identifying single-parity-check (SPC), repetition (REP) and REP-SPC nodes in the decoding tree of polar codes and implementing their fast decoders \cite{FastPolarDecoders}.

The key idea behind the above-mentioned strategies is to increase the decoding speed by reducing the decoding depth and implementing fast parallel decoders for some particular frozen-bit sequences. As such, these schemes work only on specific codes. In particular, the decoding tree of a given rate and length polar code is first constructed, and then the afore-mentioned nodes are identified and implemented in hardware. Changing the code length or rate will necessitate re-identification of these nodes. As such, these schemes are not suitable for variable-rate and/or variable-length polar codes.

The new radio access technology will use a variable rate and length polar code for downlink control information to fully utilize the physical resources \cite{HuaweiPolarCodes,QualcommLDPCCodes,EricssonPolarCode}. As such, implementing fast polar decoders that work for any rate and length is of practical interest. One such decoder, that reduces the decoding depth by one was proposed in \cite{TwoBitParallelDecoder}. In particular, the authors proposed to decode two bits in parallel by implementing four decoders corresponding to all frozen-bit sequences. The same authors then extended this concept to larger power of 2 block sizes in \cite{MultiBitParallelDecWithoutLLR,MultiBitParallelDecoder}, but the extension results in huge hardware cost as the number of all frozen-bit sequences grows exponentially with the block size. Secondly, their decoding methodology does not identify and utilize the code structure for different frozen-bit sequences to reduce the hardware area or computational complexity.

In this paper, we present fast decoders for variable-rate and/or variable-length polar codes. In particular, we borrow the idea presented in \cite{TwoBitParallelDecoder,MultiBitParallelDecWithoutLLR,MultiBitParallelDecoder} of implementing $R$-bit ($R=2,4,8,16$) parallel decoders at the last stage. But, unlike \cite{TwoBitParallelDecoder,MultiBitParallelDecWithoutLLR,MultiBitParallelDecoder}, we do not implement $2^R$ parallel decoders for decoding each contiguous block of $R$ bits. Rather, we rely on a key characteristic of properly-designed polar codes, domination contiguity of the set of good bit-channel indexes \cite{SarkisSystematicPolarCodes}, to significantly reduce the required number of parallel decoders for each block. Additionally, we use the minimum distance of the polar code corresponding to each domination-contiguous set to further reduce the number of required decoders. For example, we implement only $21$ instead of $2^{16}$ parallel decoders for $R=16$, which implies that the required number of decoders is reduced by 99.97\% compared to a simple application of ideas presented in \cite{MultiBitParallelDecWithoutLLR,MultiBitParallelDecoder}.

Achieving hardware-area reduction is not the only aim of our proposed decoding strategy. We also reduce the decoding complexity by relying on specific structure of the polar code corresponding to each bit-channel index set. We aim to minimize computationally-intensive operations, such as check-node operations, while ensuring that the proposed parallel decoders do not tangibly alter the performance of the SC decoder.

Unlike decoding, which is serial in nature, encoding of polar codes can be done in parallel. In fact, the seminal work on polar codes presented a fully-parallel encoding architecture for non-systematic polar codes \cite{PolarCodesSeminalWork}. Although very high encoding speed can be achieved by implementing a fully-parallel encoder, such an implementation is highly disadvantaged due to large memory size and number of XOR gates, especially for long polar codes. As such, folded or serial implementations of polar-code encoders are implemented to reduce the hardware area \cite{PartiallyParallelPolarEncoder}. Moreover, systematic polar codes, as proposed in \cite{ArikanSystPolarCodes}, are serial by nature.

Like decoding, the serial implementation of polar encoder results in higher encoding latency. Similar to our proposed decoding strategy, we can increase the encoding speed by implementing $R$-bit parallel encoders at the last stage. Such an implementation is particularly helpful for flexible-rate and/or flexible-length systematic polar codes as they are non-trivial to be parallelized due to their bidirectional information transfer \cite{ArikanSystPolarCodes,FastPolarDecoders,SarkisSystematicPolarCodes}.

Our main contributions are summarized in the following.
\begin{enumerate}
    \item We present fast parallel decoders for variable-rate and variable-length polar codes. In particular, we detail 9 (instead of $2^8$) decoders for 8-bit parallel decoding and 21 (instead of $2^{16}$) decoders for 16-bit parallel decoding of polar codes. These decoders accommodate all frozen-bit sequences that can occur in a block of 8 or 16 bits.

    \item Secondly, our scheme improves the encoding speed of serially-implemented variable-rate or variable-length polar codes. Our scheme is particularly useful for flexible-rate or flexible-length systematic polar codes as their encoding is hard to be parallelized.
\end{enumerate}

In the following, we first provide a background on polar codes in Section \ref{sec:PolarCodesIntro}, primarily to establish some notations and explain the challenges associate with variable-rate polar codes. We also review the domination contiguity of the set of good bit-channel indexes in Section \ref{sec:PolarCodesIntro}. We then present our proposed $R$-bit parallel encoder/decoders in Section \ref{sec:ProposedEncDecStructure}. Afterwards, we present some numerical results for corroboration of our proposed scheme, which will be followed by concluding remarks in Section \ref{sec:Conclusion}.

Although we will be focussing mainly on the systematic polar codes (due to their superior performance and difficulty in parallelization \cite{ArikanSystPolarCodes,SarkisSystematicPolarCodes}), similar results and conclusions can be drawn for non-systematic polar codes with little or no modifications.

\section{Background}
\label{sec:PolarCodesIntro}

\subsection{Encoding Polar Codes}

Polar codes defined on the binary field, $\mathbb{F}_2$, are block codes, which can be mathematically described as
\begin{IEEEeqnarray}{c}
\mathbf{x} = \mathbf{u} \mathbf{G}_N,\IEEEeqnarraynumspace
\end{IEEEeqnarray}
where $\mathbf{x}\in \mathbb{F}_2^{N}$ is the codeword of length $N=2^n$, $\mathbf{u} \in \mathbb{F}_2^{N}$ is the input vector comprising information and frozen bits, and $\mathbf{G}_N \in \mathbb{F}_2^{N \times N}$ denotes the generator matrix. The matrix $\mathbf{G}_N$ for non-reversed polar codes\footnote{For the sake of exposition, we consider only non-reversed polar codes as similar conclusions can be drawn for reversed-polar codes with proper permutation.} is $\mathbf{G}_N = \mathbf{F}^{\otimes n}$, where $\mathbf{F}^{\otimes n}$ is the $n$th tensor power of $\mathbf{F}$ defined as
\begin{IEEEeqnarray}{c}
\label{Eq:GenMatrix}
\mathbf{F}^{\otimes n} = \begin{bmatrix}\mathbf{F}^{\otimes (n-1)} & \mathbf{O} \\ \mathbf{F}^{\otimes (n-1)} & \mathbf{F}^{\otimes (n-1)}\end{bmatrix},\IEEEeqnarraynumspace
\end{IEEEeqnarray}
with $\mathbf{F}^{\otimes 0} = 1$. Denoting the left and right halves of $\mathbf{u}$ ($\mathbf{x}$) by $\mathbf{u}_0$ and $\mathbf{u}_1$ ($\mathbf{x}_0$ and $\mathbf{x}_1$), respectively, \eqref{Eq:GenMatrix} implies $\mathbf{x}_1 = \mathbf{u}_1 \mathbf{G}_{N/2}$, and $\mathbf{x}_0 = \mathbf{u}_0 \mathbf{G}_{N/2} + \mathbf{x}_1$. Consequently, a message vector of length $N$ can be encoded by encoding two message vectors of length $N/2$ each. Repeating this process $n$ times results in encoding $N$ message bits individually. As such, polar codes have a low encoding complexity of $\mathcal{O}\left(N \log_2(N)\right)$.

\subsection{Constructing Polar Codes}

Polar codes rely on the phenomenon of channel polarization, which constructs $N$ polarized bit channels out of $N$ independent copies of a given binary memoryless channel, $W$. In particular, $N$ copies of $W$ are combined and then split to a set of $N$ binary-input channels, $W_{N}^{(i)}$, where $i=0,1,\cdots,N-1$, such that the symmetric capacity of $W_{N}^{(i)}$ tends towards either $0$ or $1$ as $N$ becomes large. Bit channels having near-unity symmetric capacity are identified as `good' channels, whereas the others are classified as `bad' channels and are frozen to zero \cite{PolarCodesSeminalWork}. Mathematically, denoting the set of `good' and `bad' bit-channel indexes by $A$ and $A^c$, respectively, $\mathbf{u}_{A^c}=(u_i : i \in A^c)=\mathbf{0}$.

\subsection{Decoding Polar Codes}

\begin{figure}[!htb]
  \centering
  \includegraphics[width=\columnwidth]{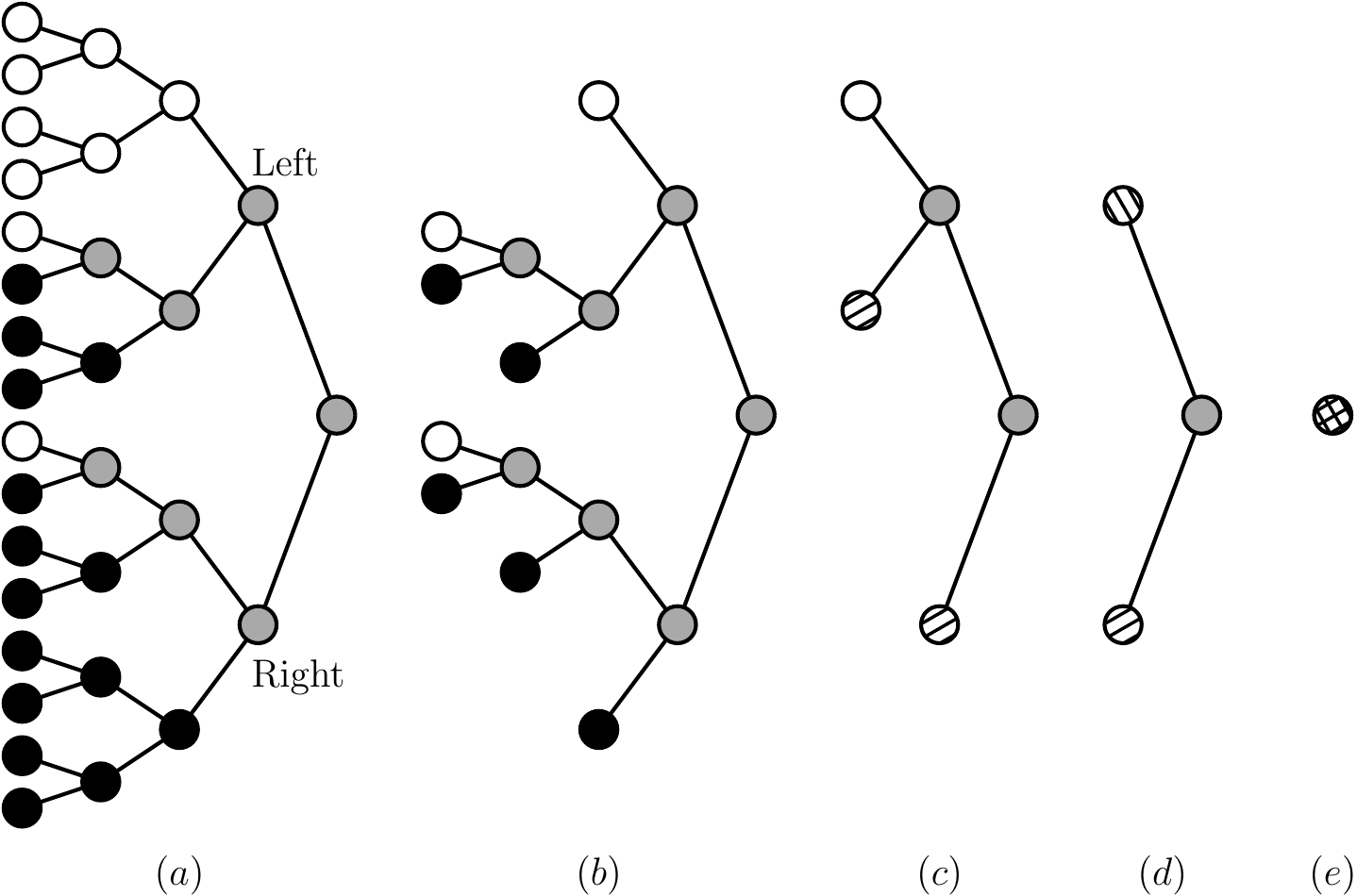}\\
  \caption{Decoding trees corresponding to the (a) SC, (b) SSC, (c) fast-SSC, (d) proposed ($R=8$), and (e) proposed ($R=16$) decoding algorithms.}\label{Fig:DecodingTrees}
\end{figure}

Since a polar code can be encoded recursively, it can be represented by a binary tree, where each node represents a codeword \cite{SimplifiedSCD,FastPolarDecoders}. Fig. \ref{Fig:DecodingTrees} (a) shows such a tree corresponding to a polar code of length 16, where the white and black leaves correspond to frozen and information bits, respectively.

We explain different decoding algorithms using the tree representation as follows. In the SC decoder \cite{PolarCodesSeminalWork}, the root node receives $\mathbf{y}$, the channel log-likelihood ratios (LLR). Denoting the left and right halves of $\mathbf{y}$ by $\mathbf{y}_0$ and $\mathbf{y}_1$, respectively, the root node sends the outputs of check-node operations between $\mathbf{y}_0$ and $\mathbf{y}_1$ to its left child. Mathematically, the left child receives $\widetilde{\mathbf{y}}_0 = \mathbf{y}_0 \boxast \mathbf{y}_1$, a vector of $N/2$ real numbers whose $i$th element, $\widetilde{y}_{0i}$, is computed as
\begin{IEEEeqnarray}{c}
  \widetilde{y}_{0i} = y_{0i} \boxast y_{1i} = 2 \tanh^{-1}\left[\tanh\left(\frac{y_{0i}}{2}\right)\tanh\left(\frac{y_{1i}}{2}\right)\right], \IEEEeqnarraynumspace
\end{IEEEeqnarray}
where $y_{0i}$ and $y_{1i}$ are the $i$th element of $\mathbf{y}_{0}$ and $\mathbf{y}_{1}$, respectively.

Afterwards, the left child performs decoding on $\widetilde{\mathbf{y}}_0$ (which we will explain later) and returns a binary vector, $\widetilde{\mathbf{x}}_0$, to the root node. The root node then computes $\widetilde{\mathbf{y}}_1$ and sends it to the right child. The $i$th element of $\widetilde{\mathbf{y}}_1$, $\widetilde{y}_{1i}$, is computed as
\begin{IEEEeqnarray}{c}
  \widetilde{y}_{1i} = \left(1-2 \widetilde{x}_{0i}\right)y_{0i} + {y}_{1i}, \IEEEeqnarraynumspace
\end{IEEEeqnarray}
where $\widetilde{x}_{0i}$ is the $i$th element of $\widetilde{\mathbf{x}}_0$. The right child then performs decoding similar to the left child and returns a binary vector, $\widetilde{\mathbf{x}}_1$, to the root node. The root node then returns an estimate of the codeword based on $\widetilde{\mathbf{x}}_0$ and $\widetilde{\mathbf{x}}_1$ as
\begin{IEEEeqnarray}{c}
  \widehat{\mathbf{x}} = \begin{bmatrix} \widetilde{\mathbf{x}}_0 +  \widetilde{\mathbf{x}}_1 & \widetilde{\mathbf{x}}_1\end{bmatrix}, \IEEEeqnarraynumspace
\end{IEEEeqnarray}
where the vector addition is performed in $\mathbb{F}_2^{N/2}$.

For non-systematic polar codes, the left and right children also return $\widehat{\mathbf{u}}_0$ and $\widehat{\mathbf{u}}_1$, respectively, to the root node. These binary vectors are estimates of the left and right halves of the input, $\mathbf{u}$. The root node, in addition to computing $\widehat{\mathbf{x}}$, computes an estimate of $\mathbf{u}$ as $\widehat{\mathbf{u}} = \begin{bmatrix}\widehat{\mathbf{u}}_0 & \widehat{\mathbf{u}}_1\end{bmatrix}$.

Since each child of the root node can be considered as the root node of a subtree, each child performs exactly the same operation of its parent node. This process continues until the leaf nodes receive real-valued messages from their parents. Since leaf nodes do not have any child, they either send 0 or hard-decision estimates based on the received LLRs to their parents depending on the frozen-bit sequence.

The decoding latency of the SC decoder depends on the computation time of the check-node operation and the decoding-tree depth. The SSC decoder \cite{SimplifiedSCD} improves the latency by identifying and removing descendants of rate-0 and rate-1 nodes in the code tree as shown in Fig. \ref{Fig:DecodingTrees} (b). Here, instead of traversing (which involves performing check-node operations) the subtree rooted in a rate-0 node, an all-zero vector is sent to its parent node. Similarly, a rate-1 node computes and sends the binary vector(s) to its parent node without traversing its descendants.

The fast-SSC decoder \cite{FastPolarDecoders} further prunes the decoding tree by identifying the SPC, REP and REP-SPC nodes in the tree. For example, in Fig. \ref{Fig:DecodingTrees} (c), the subtrees of two SPC nodes are removed from the tree resulting in the decoding depth of only two.

\begin{figure}[!htb]
  \centering
  \includegraphics[width=0.8\columnwidth]{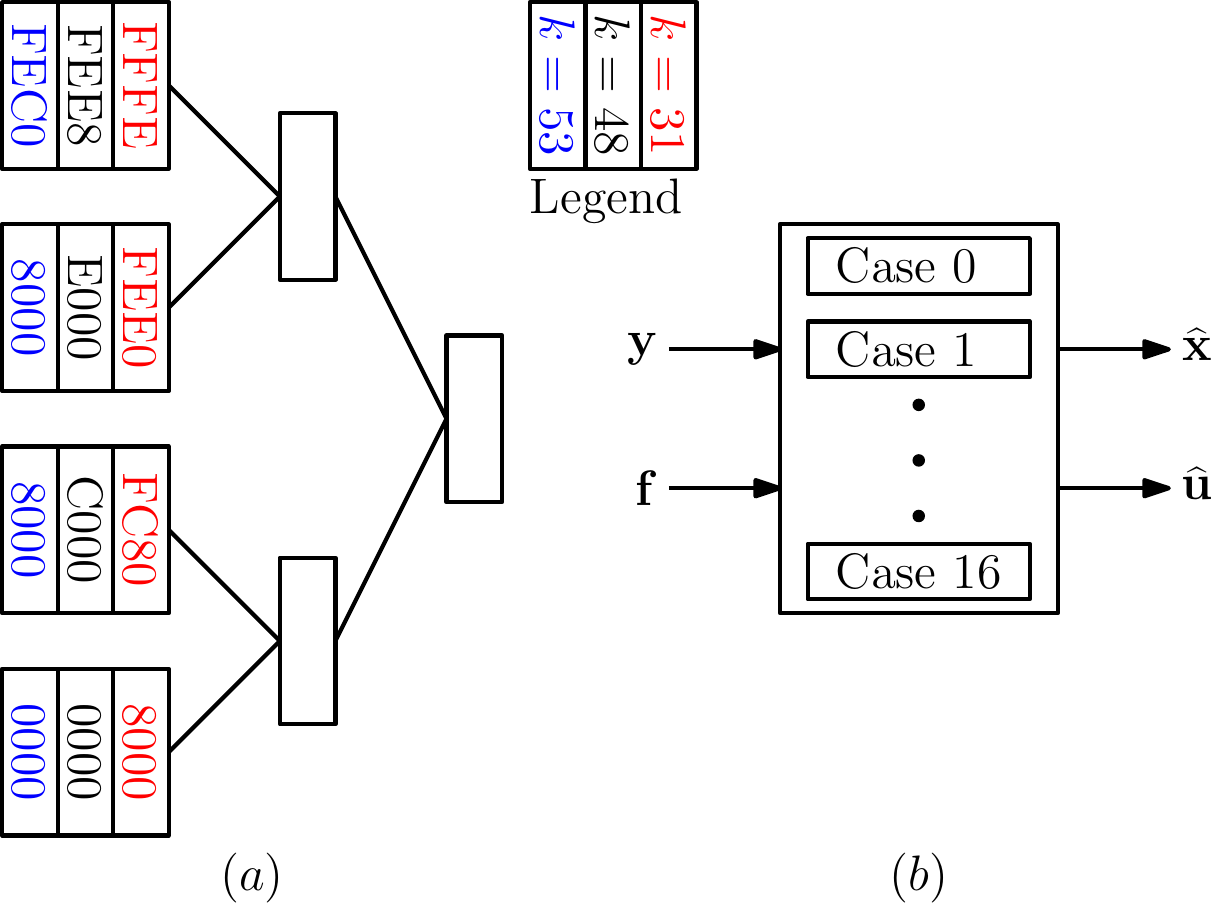}\\
  \caption{An illustration of (a) a code tree corresponding to length-64 polar codes with code rates $k/64$, where $k=31, 48,$ and $53$, and (b) the proposed 16-bit parallel decoder.}\label{Fig:N64R16Illustration}
\end{figure}

As clear from Fig. \ref{Fig:DecodingTrees}, both the SSC and fast-SSC decoders require identification of some special nodes based on the frozen-bit sequence, which can be used to eliminate corresponding subtrees in the code tree. In polar codes with flexible rate or length, the frozen-bit sequence changes with the code rate and length (see Fig. \ref{Fig:N64R16Illustration}). Hence, these algorithms are not directly applicable.

For variable-rate or variable-length polar codes, $R$-bit parallel decoders as proposed in \cite{TwoBitParallelDecoder,MultiBitParallelDecWithoutLLR,MultiBitParallelDecoder} can be used to improve the decoding speed, where $R=2,4,8, \cdots$. Fig. \ref{Fig:N64R16Illustration} shows such a case where a $16$-bit parallel decoder is implemented to decode a variable-rate polar of length $N=64$. Here, the frozen-bit sequence is represented in hexadecimal notation. For example, when the frozen-bit sequence is FFFE, all bits except the last one are frozen to zero in a block of 16 bits.

Observe that, unlike Fig. \ref{Fig:DecodingTrees} (b) and (c), the code-tree structure remains the same regardless of the code rate. Secondly, the code tree is also complete, i.e., all nodes except the leaf nodes have two children. A similar observation can be made about Fig. \ref{Fig:DecodingTrees} (d) and (e), where 8-bit and 16-bit parallel decoders are used to improve the decoding speed, respectively.

The $R$-bit parallel decoders must be able to decode all the frozen-bit sequences that can occur in a block of $R$ bits. One way to satisfy this requirement is to implement $2^R$ parallel sub-decoders and select the appropriate one corresponding to the frozen bit sequence as proposed in \cite{TwoBitParallelDecoder,MultiBitParallelDecWithoutLLR,MultiBitParallelDecoder}. Unfortunately, this solution becomes impractical even for small values of $R$. However, we can reduce the required number of decoders by using a key characteristic of properly-designed polar codes, domination contiguity, as described below.

\subsection{Domination Contiguity}
\label{sec:dominationContiguity}

Let $\BraceTwo{N}$ denote the set $\{0,1,\cdots,N-1\}$, where $N=2^n$, and $n$ is a positive integer. For $i \in \BraceTwo{N}$, we use $\langle i\rangle_2$ to represent the $n$-bit binary representation of $i$, i.e., $\langle i\rangle_2 = i_{n} i_{n-1} \cdots i_1$, where $i_n$ is the most-significant bit. For $i,j \in \BraceTwo{N}$, $i$ is said to binary dominate $j$, denoted by $i \succeq j$, if and only if $i_k \ge j_k$ for all $1 \le k \le n$, where $i_k$ and $j_k$ are the $k$th least-significant bits in the binary representations of $i$ and $j$, respectively. A set, $S \subseteq \BraceTwo{N}$, is domination contiguous if $h,j \in S$ and $h \succeq i \succeq j$ implies $i \in S$.

For a properly-designed polar code, the set of good bit-channel indexes, $A$, must be domination contiguous \cite{SarkisSystematicPolarCodes}. Since not all subsets of $\BraceTwo{N}$ are domination contiguous, the cardinality of the set of all possible frozen-bit sequences of a properly-designed polar code is less than $2^N$. In the following, we will explain how this key characteristic can be used to significantly reduce the number of required decoders from $2^R$ for $R$-bit parallel decoders.

\section{Proposed Encoder/Decoder Structure}
\label{sec:ProposedEncDecStructure}

The proposed encoders/decoders rely on parallel encoding/decoding of $R$ bit channels simultaneously, where $R=2^t$, and $t$ is a positive integer. Specifically, we divide $N$-bit $\mathbf{u}$ into $N/R$ consecutive groups each containing $R$ bits, i.e., $\mathbf{u}=[\mathbf{u}_0 \cdots \mathbf{u}_{N/R-1}]$. Then we encode/decode $R$ bits in each group simultaneously to reduce the encoding/decoding depth of the code tree by $t$ levels.

Quite intuitively, the greater the value of $R$, the faster the encoder/decoder will be. On the downside, we require more hardware to implement all the encoders/decoders for a block size of $R$. Theorem \ref{Theorem:BlockContiguity} describes inheritance of domination contiguity of $A$ to the domination contiguity in each block of $R$ bits, which will be used later to optimize the number of required encoders/decoders for an $R$-bit parallel decoder.

\begin{theorem}
\label{Theorem:BlockContiguity}
Let $\tilde{A}_i$ denote the set of good bit-channel indexes corresponding to $\mathbf{u}_i$, $i=0,1,\cdots,N/R-1$. For a properly-designed polar code, $\tilde{A}_i$ is domination contiguous.
\end{theorem}

\begin{IEEEproof}
Observe that all elements of each $\tilde{A}_i$ have the same $n-t$ most-significant bits, where $n=\log_2(N)$, and $t=\log_2(R)$. Further, the $n-t$ most-significant bits corresponding to each $\tilde{A}_i$ differ in at least one bit from that of $\tilde{A}_j$, where $j=0,1,\cdots,N/R-1$, and $j \neq i$. Domination contiguity of $A$ implies that if $h,j \in \tilde{A}_i$ and $h \succeq k \succeq j$ then $k \in A$. But $h$ and $j$ have the same $n-t$ most-significant bits. As such, $h \succeq k \succeq j$ implies that $k$ also has the same $n-t$ most-significant bits. Therefore, $k \in \tilde{A}_i$.
\end{IEEEproof}

An immediate consequence of the domination contiguity of $\tilde{A}_i$'s is that the number of required encoders/decoders can be reduced from $2^R$. The following theorem shows that the number of all domination-contiguous sets becomes significantly less than $2^R$ as $R$ grows to infinity.

\begin{theorem}
\label{Theroem:NoOfDominContig}
Let $\mathcal{U}_R^i = \{\tilde{A}_i\}$ denote a universal set containing all the good bit-channel indexes of the $i$th $R$-bit block, $\mathbf{u}_i$, for $i=0,1,\cdots,N/R-1$. Denoting the cardinality of a set $S$ by $|S|$, the ratio $|\mathcal{U}_R^i|/2^R$ is a decreasing function of $R$, and in the limit as $R$ goes to infinity, the ratio $|\mathcal{U}_R^i|/2^R$ goes to zero.
\end{theorem}

\begin{IEEEproof}
Without loss of generality, we consider the first $R$-bit block ($i=0$). Furthermore, we let $P_R = |\mathcal{U}_R^0|$ and split the elements of any $\tilde{A}_0 \in \mathcal{U}_R^0$ into two sets, $S_0$ and $S_1$, where $S_0$ contains those elements of $\tilde{A}_0$ that are less than $R/2$, and $S_1$ contains the remaining.

Theorem \ref{Theorem:BlockContiguity} asserts that the domination contiguity of $\tilde{A}_0$ begets domination contiguity of $S_0$ and $S_1$ . Consequently, $\mathcal{U}_R^0 \subseteq \mathcal{B}$, where $\mathcal{B}=\{B:B=B_0 \cup B_1\}$, $B_0 \in \mathcal{U}_{R/2}^0$, $B_1 = \{b_i: (b_i-R/2) \in \tilde{B}\}$, and  $\tilde{B}\in \mathcal{U}_{R/2}^0$. Clearly, $|\mathcal{B}|=P_{R/2}^2$, and as such, $|\mathcal{U}_R^0| = P_R \le P_{R/2}^2$.

Moreover, if $\tilde{A}_0 \neq \{\}$ then $R-1 \in \tilde{A}_0$ \cite{FastPolarDecoders}. Equivalently, $\tilde{A}_0 \neq \{\}$ implies $S_1 \neq \{\}$. Conversely, if $S_1=\{\}$ then $\tilde{A}_0=\{\}$ and $S_0=\{\}$. But $\mathcal{B}$ contains those sets which correspond to $S_1 = \{\}$ and $S_0 \neq \{\}$. Therefore, $\mathcal{U}_R^0 \subset \mathcal{B}$, and $P_R < P_{R/2}^2$.

Lastly, we define a sequence $a_t = P_{R}/2^{R} = P_{2^t}/2^{2^t}$, where $t=\log_2(R)$, and $t=1,2,\cdots$. Using the fact that $0 < a_t < 1$ and $a_{t+1} < a_{t}^2$, we get $\lim_{t \rightarrow \infty}{a_t}= \lim_{R \rightarrow \infty}{\frac{P_R}{2^R}} = 0$.
\end{IEEEproof}

Theorem \ref{Theroem:NoOfDominContig} shows that the ratio $P_R/2^R$ is a decreasing function of $R$. But it does not imply that $P_R$ does not increase rapidly.  In fact, $P_R$ can be shown to be equal to $2$, $3$, $6$, $20$, $168$, and $7581$ for $R = 1,2,4,8,16,$ and $32$, respectively. Clearly, implementing $P_R$ parallel decoders to increase the decoding speed becomes impractical even for moderate block lengths. Fortunately, we can reduce the number of required decoders further by eliminating some frozen-bit sequences depending on the minimum distance of the corresponding code and the position of frozen bits in the sequence as explained below. Note that the minimum required number of encoders/decoders corresponding to each block of $R$ bits is $R+1$ because the number of frozen bits in a block of $R$ bits can vary from $0$ to $R$. Hence, at least $R+1$ encoders/decoders are needed. A solution based on $R+1$ encoders/decoders, if exists, has minimized the hardware area.

In the following, we present encoders for all the $20$ cases corresponding to $\tilde{A}_i$ for $R=8$. Later, we reduce this number to the minimum, i.e. 9, based on the position of frozen bits and the minimum distance of the polar code corresponding to each frozen-bit sequence. To do so, we define the notion of max-frozen set, which will be used to eliminate some candidate sets.

Let $\pi_n : \BraceTwo{N} \rightarrow \BraceTwo{N}$  denote a bit-permutation function that maps $j \in \BraceTwo{N}$ to $\pi_n(j)$ such that the bits in $\langle \pi_n(i) \rangle_2$ are a permutation of the bits in $\langle i \rangle_2$. Further, for $\mathcal{D} \subseteq \BraceTwo{N}$, we define $\pi_n\left(\mathcal{D}\right)=\{\pi_n(d): d \in \mathcal{D}\}$. For example, the bit-reversal permutation \cite{PolarCodesSeminalWork} for $N=4$ maps $j$ to $\pi_2(j)$, where $\langle \pi_2(j) \rangle_2 = j_1 j_2$, and $\langle j \rangle_2 = j_2 j_1$. Likewise, if $\mathcal{D} = \{0,2,3\}$ then $\pi_2\left(\mathcal{D}\right) = \{0,1,3\}$.

\begin{lemma}
\label{Lemma:ConjugateContiguity}
If $\mathcal{D}$ is domination contiguous then so is $\pi_n\left(\mathcal{D}\right)$.
\end{lemma}

\begin{IEEEproof}
Let $\pi_n^{-1}(\cdot)$ be the inverse permutation function, i.e., $\pi_n^{-1}(\pi_n(j))=j$. Further, let $h,j \in \pi_n\left(\mathcal{D}\right)$ and $h \succeq k \succeq j$. As such, $\pi_n^{-1}(h), \pi_n^{-1}(j) \in \mathcal{D}$. Also, we observe that binary domination is invariant to bit permutations, i.e., $h \succeq k \succeq j$ implies $\pi_n^{-1}(h) \succeq \pi_n^{-1}(k) \succeq \pi_n^{-1}(j)$. Since domination contiguity of $\mathcal{D}$ implies $\pi_n^{-1}(k) \in \mathcal{D}$, $k \in \pi_t \left(\mathcal{D}\right)$. Therefore, $\pi_n\left(\mathcal{D}\right)$ is domination contiguous.
\end{IEEEproof}

Let $\mathbf{P}_N = [\mathbf{e}_{\pi_n(0)} \cdots \mathbf{e}_{\pi_n(N-1)}]$ denote the permutation matrix corresponding to the bit-permutation function $\pi_n(\cdot)$. Here, $\mathbf{e}_l$ denotes a column vector whose $l$th element is 1 and the remaining $N-1$ elements are 0. It was shown in \cite{DualPolarCode,PermutedSCD} that $\mathbf{P}_N \mathbf{G}_N = \mathbf{G}_N \mathbf{P}_N$. Therefore, $\mathbf{u} \mathbf{P}_N \mathbf{G}_N = \mathbf{x} \mathbf{P}_N$. Denoting $\mathbf{u} \mathbf{P}_N = \mathbf{u}_{\pi_{n}}$ and $\mathbf{x} \mathbf{P}_N = \mathbf{x}_{\pi_{n}}$, we get $\mathbf{u}_{\pi_{n}} \mathbf{G}_N = \mathbf{x}_{\pi_{n}}$. Consequently, if $\mathbf{x}$ is a polar code for input $\mathbf{u}$ then permuting the bit positions of $\mathbf{u}$ permutes $\mathbf{x}$ in the same manner. We refer these codes to be the \emph{conjugates} of the original code. Observe that the set of good bit-channel indexes of a conjugate polar code is $\pi_n \left( A \right)$.

Lemma \ref{Lemma:ConjugateContiguity} confirms that, similar to $A$, $\pi_n \left( A \right)$ is domination contiguous. Further, when $\pi_n \left( A \right)$  differs $A$, the SC decoding shows worse performance \cite{PermutedSCD}. However, the SC decoder can be modified to decode in the permuted order \cite{DualPolarCode,PermutedSCD} to achieve the same performance. Consequently, for a given decoding order, only one set of good bit-channel indexes will show the best performance. Quite intuitively, the set which results in early `decoding' of the most frozen bits will outperform others \cite[Section 7.4.3]{PolarCodePermutationThesis}. We call this set the \emph{max-frozen set}. As such, for a given decoding order, the number of required decoders can be reduced by implementing decoders only for the max-frozen sets and ignoring their distinct bit-permuted sets.

In the following, we present encoders and decoders for $R=8$. For the sake of clarity, we will drop the subscript $i$ from $\tilde{A}_i$. Also, the set of frozen bit-channel indexes will be represented by $\tilde{A}^c$. Lastly, we will consider the natural-order decoding of polar codes to eliminate all the bit-permuted sets of the max-frozen set.

\subsection{Block Size 8 Encoders/Decoders}
\label{sec:R8EncDec}

\begin{table*}[!htb]
\renewcommand{\arraystretch}{1.3}
\caption{All possible frozen-bit sequences and the corresponding outputs for $R=8$.}\label{AllCasesR8}
\centering
\begin{tabular}{c||l|l|l|l|l}
  \hline
  $k$ & $\mathbf{f}$ (Binary) & $\mathbf{f}$ (Hex)& $\tilde{A}$ & \textbf{Output}, $\mathbf{x}$  & $d_{\min}$\\
  \hline
  0 & $1,1,1,1,1,1,1,1$ & FF &\{\}& $0,0,0,0,0,0,0,0$ & - \\ \hline
  1 & $1,1,1,1,1,1,1,0$ & FE &\{7\}& $\x{7},\x{7},\x{7},\x{7},\x{7},\x{7},\x{7},\x{7}$ & 8 \\ \hline
    & $1,1,1,1,1,1,0,0$ & FC &\{6,7\}& $\x{6},\x{7},\x{6},\x{7},\x{6},\x{7},\x{6},\x{7}$ & 4 \\ \cline{2-6}
  2 & $1,1,1,1,1,0,1,0$ & FA &\{5,7\}& $\x{5},\x{5},\x{7},\x{7},\x{5},\x{5},\x{7},\x{7}$ & 4 \\ \cline{2-6}
    & $1,1,1,0,1,1,1,0$ & EE &\{3,7\}& $\x{3},\x{3},\x{3},\x{3},\x{7},\x{7},\x{7},\x{7}$ & 4 \\ \hline
    & $1,1,1,1,1,0,0,0$ & F8 &\{5,6,7\}& $\x{567},\x{5},\x{6},\x{7},\x{567},\x{5},\x{6},\x{7}$ & 4 \\ \cline{2-6}
  3 & $1,1,1,0,1,1,0,0$ & EC &\{3,6,7\}& $\x{367},\x{3},\x{367},\x{3},\x{6},\x{7},\x{6},\x{7}$ & 4 \\ \cline{2-6}
    & $1,1,1,0,1,0,1,0$ & EA &\{3,5,7\}& $\x{357},\x{357},\x{3},\x{3},\x{5},\x{5},\x{7},\x{7}$ & 4 \\ \hline
  \multirow{4}{*}{4}
    & $1,1,1,0,1,0,0,0$ & E8 &\{3,5,6,7\}& $\x{356},\x{357},\x{367},\x{3},\x{567},\x{5},\x{6},\x{7}$ & 4 \\ \cline{2-6}
    & $1,1,1,1,0,0,0,0$ & F0 &\{4,5,6,7\}& $\x{4},\x{5},\x{6},\x{7},\x{4},\x{5},\x{6},\x{7}$ & 2 \\ \cline{2-6}
    & $1,1,0,0,1,1,0,0$ & CC &\{2,3,6,7\}& $\x{2},\x{3},\x{2},\x{3},\x{6},\x{7},\x{6},\x{7}$ & 2 \\ \cline{2-6}
    & $1,0,1,0,1,0,1,0$ & AA &\{1,3,5,7\}& $\x{1},\x{1},\x{3},\x{3},\x{5},\x{5},\x{7},\x{7}$ & 2 \\ \hline

    & $1,1,1,0,0,0,0,0$ & E0 &\{3,4,5,6,7\}& $\x{347},\x{357},\x{367},\x{3},\x{4},\x{5},\x{6},\x{7}$ & 2 \\ \cline{2-6}
  5 & $1,1,0,0,1,0,0,0$ & C8 &\{2,3,5,6,7\}& $\x{257},\x{357},\x{2},\x{3},\x{567},\x{5},\x{6},\x{7}$ & 2 \\ \cline{2-6}
    & $1,0,1,0,1,0,0,0$ & A8 &\{1,3,5,6,7\}& $\x{167},\x{1},\x{367},\x{3},\x{567},\x{5},\x{6},\x{7}$ & 2 \\ \hline
    & $1,1,0,0,0,0,0,0$ & C0 &\{2,3,4,5,6,7\}& $\x{246},\x{357},\x{2},\x{3},\x{4},\x{5},\x{6},\x{7}$ & 2 \\ \cline{2-6}
  6 & $1,0,1,0,0,0,0,0$ & A0 &\{1,3,4,5,6,7\}& $\x{145},\x{1},\x{367},\x{3},\x{4},\x{5},\x{6},\x{7}$ & 2 \\ \cline{2-6}
    & $1,0,0,0,1,0,0,0$ & 88 &\{1,2,3,5,6,7\}& $\x{123},\x{1},\x{2},\x{3},\x{567},\x{5},\x{6},\x{7}$ & 2 \\ \hline
  7 & $1,0,0,0,0,0,0,0$ & 80 &\{1,2,3,4,5,6,7\}& $\x{1234567},\x{1},\x{2},\x{3},\x{4},\x{5},\x{6},\x{7}$ & 2 \\ \hline
  8 & $0,0,0,0,0,0,0,0$ & 00 &\{0,1,2,3,4,5,6,7\}& $\x{0},\x{1},\x{2},\x{3},\x{4},\x{5},\x{6},\x{7}$ & 1 \\ \hline
\end{tabular}
\end{table*}

Table \ref{AllCasesR8} enlists all the 20 frozen-bit sequences, $\mathbf{f}$, that can occur in a block of 8 consecutive bit channels of a properly-designed polar code. The $i$th component of $\mathbf{f}$ is 1 if the $i$th bit channel is frozen and is 0 otherwise. These sequences are grouped into 9 different cases depending on the number of information bits in the block of 8 bits. The corresponding set of good bit-channel indexes are also tabulated. Observe that all the sets are domination contiguous. Lastly, the systematic polar code, $\mathbf{x}$, corresponding to each frozen-bit sequence is also mentioned along with its minimum distance, $d_{\min}$. As we explain below, the code structure and its corresponding frozen-bit sequence and $d_{\min}$ will be used to further reduce the number of possible cases from 20 to 9. For the sake of brevity, we have used the notation $x_{abc}$ to denote $x_a + x_b + x_c$.

In the following, we discuss each individual case and explain why a particular frozen-bit sequence is kept in each case.

\subsubsection{Case 0}
This case corresponds the rate-0 node introduced in \cite{SimplifiedSCD}, and the optimal decoder assigns an all-zero vector to the output.

\subsubsection{Case 1}

This is an $(8,1)$ repetition code, and the optimal maximum-likelihood (ML) decoder will add the LLRs of all the channel outputs and perform threshold detection on the sum \cite{MLSoftDecodBlockCode}. The same decoder is used in \cite{FastPolarDecoders}, where they have outlined some low-latency decoding strategies for improving the decoding speed.

\subsubsection{Case 2}

All the three cases are $(8,2)$ repetition codes and are conjugates of one another. For the SC decoder with natural-order decoding, the first case is the max-frozen set. Consequently, other cases will not occur. The optimal decoder, like Case 1, will add the LLRs of four outputs to estimate $x_7$ and other four LLRs to estimate $x_6$.

\subsubsection{Case 3}

These codes are concatenated (8,4) repetition and (4,3) single parity-check (SPC) codes and are conjugates of one another. Since $\tilde{A}=\{5,6,7\}$ is the max-frozen set, only the first case will occur in practice. The decoding can be carried out by first adding the LLRs of the outputs corresponding to the same bits. As such, we are left with the LLRs of a (4,3) SPC code. The optimal ML decoder of the SPC codes, Wagner decoder \cite{WagnerDecoder}, makes hard-decision estimates of $x_i$'s and flips the least-reliable bit if the parity check is not satisfied.

\subsubsection{Case 4}

The number of good bit-channel indexes, $\tilde{A}$, is 4 when $k=4$. Amongst them, three correspond to $(8,4)$ repetition codes. As such, their $d_{\min}=2$. In only one case, $\tilde{A}=\{3,5,6,7\}$, the minimum distance turns out to be 4. Since code performance heavily depends on $d_{\min}$, only this case will occur in practice. In fact, this is an extended Hamming code \cite{ErrorControlDigCommWicker} and is equivalent to the repetition-SPC code introduced in \cite{FastPolarDecoders}. Although the optimal ML decoder for such a code can be implemented easily \cite{MLSoftDecodBlockCode}, a low-complexity decoder of this code was mentioned in \cite{FastPolarDecoders}. Furthermore, the bit-error rate (BER) performance of the code is not considerably altered by implementing the low-complexity decoder instead of the optimal decoder \cite{FastPolarDecoders}. For completeness, we briefly mention the low-complexity decoder below.

First, observe that $(x_0+x_4,x_1+x_5,x_2+x_6,x_3+x_7)$ constitute a (4,1) repetition code, while $(x_4,x_5,x_6,x_7)$ is a (4,3) SPC code. The repetition code can easily be decoded by adding the LLRs, resulting in $\widehat{x}_8$, a hard-decision estimate of $x_8 = x_3+x_7$. Afterwards, additional LLRs for $(x_4,x_5,x_6,x_7)$ are trivially computed either by keeping or switching the sign of the LLRs of $(x_0,x_1,x_2,x_3)$ depending on the value of $\widehat{x}_8$. After adding the LLRs of $(x_4,x_5,x_6,x_7)$, we are left with a (4,3) SPC code, which can be decoded by the Wagner decoder.

\subsubsection{Case 5}

\begin{figure}
  \centering
  \includegraphics[width=0.5\columnwidth]{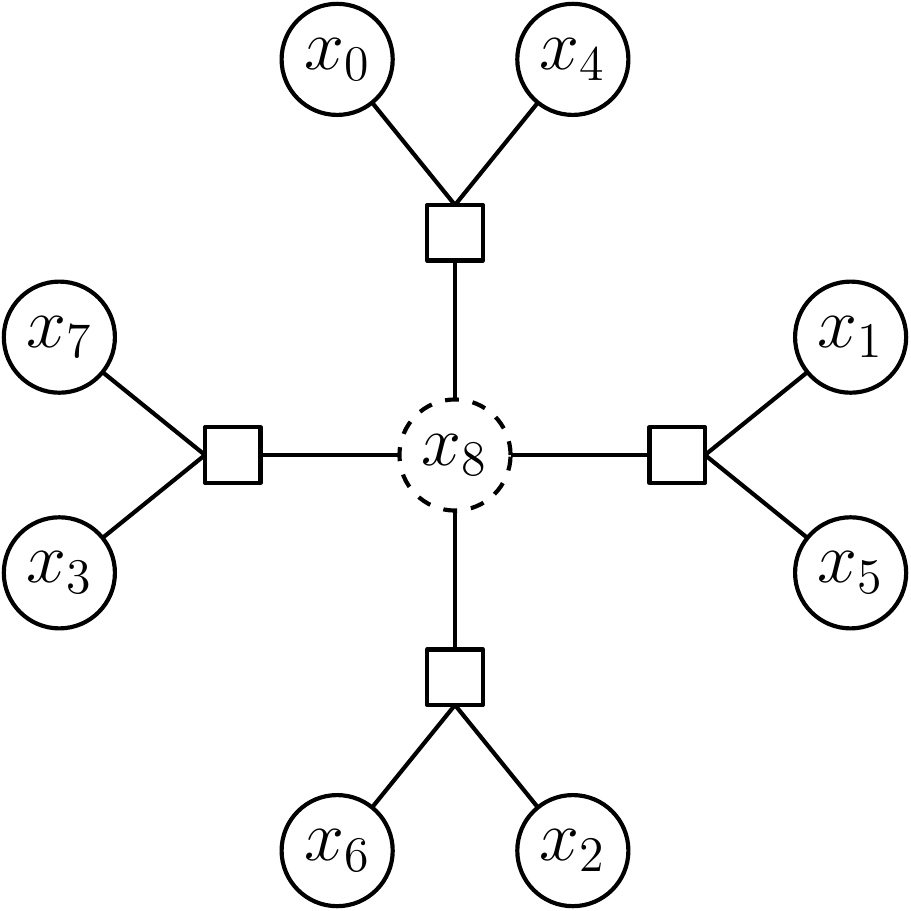}\\
  \caption{Tanner graph for the systematic-polar code corresponding to $\tilde{A}=\{3,4,5,6,7\}$.}\label{Case5TannerGraph}
\end{figure}

Observe that all the codes are conjugate of one another, and $\tilde{A}=\{3,4,5,6,7\}$ is the max-frozen set. Thus, only the first case will occur in practice. By introducing an additional node, $x_8 = x_3 + x_7$, a cycle-free Tanner graph of the code can be obtained as shown in Fig. \ref{Case5TannerGraph}. As such, a non-iterative optimal maximum-a-posteriori (MAP) decoder can be implemented \cite{CycleFreeFactorGraph}.

Like Case 4, a low-complexity sub-optimal decoder can also be implemented by first making a hard decision about $x_8$. Afterwards, depending on $\widehat{x}_8$, LLRs of $(x_0,x_1,x_2,x_3)$ can be added or subtracted to that of $(x_4,x_5,x_6,x_7)$. Hard decisions of $(x_4,x_5,x_6,x_7)$ can then be carried out, which along with $\widehat{x}_8$ can be used to find estimates of $(x_0,x_1,x_2,x_3)$. The decoding latency can be reduced by implementing two parallel decoders assuming $\widehat{x}_8$ equals 0 or 1 and selecting the appropriate output depending on the actual value of $\widehat{x}_8$. Also, as verified by the simulation results, the BER performance is not tangibly degraded by implementing the sub-optimal decoder instead of the optimal MAP decoder.

\subsubsection{Case 6}

All of the codes conjugates of one another. But the first case will occur in practice as $\tilde{A}=\{2,3,4,5,6,7\}$ is the max-frozen set. Since this code is just two (4,3) SPC codes put together, two Wagner decoders can be used to optimally decode it.

\subsubsection{Case 7}

This is an (8,7) SPC code, and the Wagner decoder can be used to optimally decode it. Note that this code is equivalent to the one corresponding to the SPC node mentioned in \cite{FastPolarDecoders}, which were introduced to increase the decoding speed, especially for high-rate polar codes.

\subsubsection{Case 8}

This case is equivalent to the rate-1 node introduced in \cite{SimplifiedSCD}, and a hard decision of the observed outputs gives the decoded output.

\begin{remark}

For non-systematic codes, exactly the same decoders can be used to find a hard-decision estimate of $\mathbf{x}$, which can be used to decode the input, $\mathbf{u}$. For example, for $\mathbf{f}=(1,1,1,1,1,1,0,0)$, $\mathbf{x}=(u_6+u_7,u_7,u_6+u_7,u_7,u_6+u_7,u_7,u_6+u_7,u_7)$, which is an (8,2) repetition code. Like in Case 2, we can find hard-decision estimates of $u_6+u_7$ and $u_7$, which can be used to find a hard-decision estimate of $u_6$.
\end{remark}

\begin{remark}
For permuted polar codes, similar conclusions can be drawn as the set of good bit-channel indexes is also domination contiguous for permuted (systematic/non-systematic) polar codes. In particular, the max-frozen sets for an individual case will be determined according to the permuted decoding order. Also, the presented low-complexity decoders can trivially be modified to decode the polar codes corresponding to the max-frozen sets.
\end{remark}

\begin{remark}

Implementing the proposed low-complexity decoders reduces the number of check-node operations significantly. In particular, the SC decoder uses 12 check-node operations for decoding 8 bits for each of the afore-mentioned cases. In our case, however, baring Case 4 and Case 5, check-node operations are not used. Furthermore, decoders of Case 4 and Case 5 use only 4 check-node operations, which are implemented in parallel. As such, compared with the SC decoder, where 12 check-node operations are used and are carried out sequentially, the proposed decoders require less check-node operations and are faster as these operations can be implemented in parallel.

\end{remark}

\subsection{Block Size 16 Encoders/Decoders}
\label{sec:R16EncDec}

Having discussed all the possible cases for $R=8$, we now consider polar codes for $R=16$. Similar to $R=8$ case, we can significantly reduce the required number of parallel encoders/decoders (from 168 to just 21). Table \ref{AllCasesR16} enlists these 21 cases along with the $d_{\min}$ of the corresponding codes. Here, we have used hexadecimal indexes for the sake of brevity. The appendix provides a detailed description of the proposed decoders for the cases tabulated in Table \ref{AllCasesR16}.

\begin{table*}[!htb]
\renewcommand{\arraystretch}{1.3}
\caption{Frozen-bit sequences and the corresponding outputs for $R=16$.}\label{AllCasesR16}
\centering
\begin{tabular}{c||l|l|c}
  \hline
  $k$ & $\mathbf{f}$ (Hex) & \textbf{Output}, $\mathbf{x}$ & $d_{\min}$\\
  \hline
  0 & FFFF & $0,0,0,0,0,0,0,0,0,0,0,0,0,0,0,0$ & -\\ \hline
  1 & FFFE & $\x{f},\x{f},\x{f},\x{f},\x{f},\x{f},\x{f},\x{f},\x{f},\x{f},\x{f},\x{f},\x{f},\x{f},\x{f},\x{f}$ & 16\\ \hline
  2 & FFFC & $\x{e},\x{f},\x{e},\x{f},\x{e},\x{f},\x{e},\x{f},\x{e},\x{f},\x{e},\x{f},\x{e},\x{f},\x{e},\x{f}$ & 8\\ \hline
  3 & FFF8 & $\x{def},\x{d},\x{e},\x{f},\x{def},\x{d},\x{e},\x{f},\x{def},\x{d},\x{e},\x{f},\x{def},\x{d},\x{e},\x{f}$
  & 8 \\ \hline
  4 & FFE8 & $\x{bde},\x{bdf},\x{bef},\x{b},\x{def},\x{d},\x{e},\x{f}, \x{bde},\x{bdf},\x{bef},\x{b},\x{def},\x{d},\x{e},\x{f}$  & 8\\ \hline
  5 & FEE8 & $\x{7bdef},\x{7bd},\x{7be},\x{7bf},\x{7de},\x{7df},\x{7ef},\x{7}, \x{bde},\x{bdf},\x{bef},\x{b},\x{def},\x{d},\x{e},\x{f}$ & 8\\ \hline
  \multirow{2}{*}{6}
    & FFC0 & $\x{ace},\x{bdf},\x{a},\x{b},\x{c},\x{d},\x{e},\x{f}, \x{ace},\x{bdf},\x{a},\x{b},\x{c},\x{d},\x{e},\x{f}$ & 4 \\ \cline{2-4}
    & FEE0 & $\x{7bc},\x{7bd},\x{7be},\x{7bf},\x{7cf},\x{7df},\x{7ef}, \x{7},\x{bcf},\x{bdf},\x{bef},\x{b},\x{c},\x{d},\x{e},\x{f}$ & 4\\ \hline
  \multirow{2}{*}{7}
    & FF80 & $\x{9abcdef},\x{9},\x{a},\x{b},\x{c},\x{d},\x{e},\x{f}, \x{9abcdef},\x{9},\x{a},\x{b},\x{c},\x{d},\x{e},\x{f}$ & 4 \\ \cline{2-4}
    & FEC0 & $\x{7acef},\x{7bd},\x{7af},\x{7bf},\x{7cf},\x{7df},\x{7ef},\x{7}, \x{ace},\x{bdf},\x{a},\x{b},\x{c},\x{d},\x{e},\x{f}$ & 4\\  \hline
  \multirow{2}{*}{8}
    & FE80 & $\x{79abcdf},\x{79f},\x{7af},\x{7bf},\x{7cf},\x{7df},\x{7ef},\x{7}, \x{9abcdef},\x{9},\x{a},\x{b},\x{c},\x{d},\x{e},\x{f}$ & 4\\ \cline{2-4}
    & FCC0 & $\x{6ac},\x{7bd},\x{6ae},\x{7bf},\x{6ce},\x{7df},\x{6},\x{7}, \x{ace},\x{bdf},\x{a},\x{b},\x{c},\x{d},\x{e},\x{f}$ & 4 \\   \hline
  9 & FC80 & $\x{69abcdf},\x{79f},\x{6ae},\x{7bf},\x{6ce},\x{7df},\x{6},\x{7}, \x{9abcdef},\x{9},\x{a},\x{b},\x{c},\x{d},\x{e},\x{f}$  & 4\\  \hline
  10 & F880 & $\x{5679abc},\x{59d},\x{6ae},\x{7bf},\x{567cdef},\x{5},\x{6},\x{7}, \x{9abcdef},\x{9},\x{a},\x{b},\x{c},\x{d},\x{e},\x{f}$ & 4 \\  \hline
  11 & E880 &  $\x{3569acf},\x{3579bdf},\x{367abef},\x{3},\x{567cdef},\x{5},\x{6},\x{7}, \x{9abcdef},\x{9},\x{a},\x{b},\x{c},\x{d},\x{e},\x{f}$ & 4\\ \hline

  \multirow{2}{*}{12}
     & E800 & $\x{3568bde},\x{3579bdf},\x{367abef},\x{3},\x{567cdef},\x{5},\x{6},\x{7}, \x{8},\x{9},\x{a},\x{b},\x{c},\x{d},\x{e},\x{f}$ & 2\\  \cline{2-4}
     & C0C0 & $\x{246},\x{357},\x{2},\x{3},\x{4},\x{5},\x{6},\x{7},\x{ace},\x{bdf},\x{a},\x{b},\x{c},\x{d},\x{e},\x{f}$ & 2 \\ \hline
  13 & E000 & $\x{3478bcf},\x{3579bdf},\x{367abef},\x{3},\x{4},\x{5},\x{6},\x{7}, \x{8},\x{9},\x{a},\x{b},\x{c},\x{d},\x{e},\x{f}$ & 2\\  \hline
  14 & C000 & $\x{2468ace},\x{3579bdf},\x{2},\x{3},\x{4},\x{5},\x{6},\x{7}, \x{8},\x{9},\x{a},\x{b},\x{c},\x{d},\x{e},\x{f}$ & 2 \\  \hline
  15 & 8000 & $\x{123456789abcdef},\x{1},\x{2},\x{3},\x{4},\x{5},\x{6},\x{7}, \x{8},\x{9},\x{a},\x{b},\x{c},\x{d},\x{e},\x{f}$ & 2 \\   \hline
  16 & 0000 & $\x{0},\x{1},\x{2},\x{3},\x{4},\x{5},\x{6},\x{7},\x{8},\x{9},\x{a},\x{b},\x{c},\x{d},\x{e},\x{f}$ & 1 \\
  \hline
\end{tabular}
\end{table*}

It is worth noting that at least 17 encoders/decoders are required for encoding/decoding flexible-rate polar codes. So only four extra encoders/decoders are needed to ensure that the polar codes designed for any rate, length and channel can be encoded/decoded. The following theorems assert that the encoders/decoders for $\mathbf{f}=\mathrm{FFC0}$, $\mathbf{f}=\mathrm{FF80}$, $\mathbf{f}=\mathrm{FCC0}$, and $\mathbf{f}=\mathrm{C0C0}$ are not required when polar codes are designed for a binary-erasure channel (BEC) or by Huawei formula \cite{HuaweiPolarCodes}.

\begin{theorem}
If polar codes are constructed for a BEC, encoders/decoders for $\mathbf{f}=\mathrm{FFC0}$, $\mathbf{f}=\mathrm{FF80}$, $\mathbf{f}=\mathrm{FCC0}$, and $\mathbf{f}=\mathrm{C0C0}$ are not required.
\end{theorem}

\begin{IEEEproof}
This assertion can be proved by noting that, regardless of the value of erasure probability, the afore-mentioned four cases do not occur when $N=16$. The result immediately follows by noting that the frozen-bit sequence for each of 16-bit block is generated for a BEC \cite{PolarCodesSeminalWork}.
\end{IEEEproof}

Recently, in 3GPP RAN1 \#87 meeting, an agreement was reached to use variable-rate polar codes for uplink control channel \cite{EricssonPolarCode,HuaweiPolarCodes}. Since polar-code design is channel dependent, and the location of frozen bits varies with the channel conditions, Huawei presented a channel-independent reliability metric for constructing polar codes \cite{HuaweiPolarCodes}. In particular, each polarized bit-channel, $W_N^{(j)}$, is assigned a reliability metric, $Q_j$, computed as
\begin{IEEEeqnarray}{c}
  Q_j = \sum_{k=1}^{n} {j_k}2^{\frac{k-1}{4}}, \IEEEeqnarraynumspace
\end{IEEEeqnarray}
where $j_k$ is the $k$th least-significant bit in the $n$-bit binary representation of $j$, i.e., $\langle j\rangle_2 = j_n j_{n-1} \cdots j_1$.

The following theorem asserts that the extra cases are not required when polar codes are constructed by Huawei formula.

\begin{theorem}
If polar codes are constructed by using Huawei formula, the four extra case ($\mathbf{f}=\mathrm{FFC0}$, $\mathbf{f}=\mathrm{FF80}$, $\mathbf{f}=\mathrm{FCC0}$, and $\mathbf{f}=\mathrm{C0C0}$) do not occur at all.
\end{theorem}

\begin{IEEEproof}
Observe that
\begin{IEEEeqnarray}{c}
  Q_j = \sum_{k=1}^{n} {j_k}2^{\frac{k-1}{4}} =  \underbrace{\sum_{k=1}^{4} {j_k}2^{\frac{k-1}{4}}}_{T_1} + \underbrace{\sum_{k=5}^{n} {j_k}2^{\frac{k-1}{4}}}_{T_2}, \IEEEeqnarraynumspace
\end{IEEEeqnarray}
Next, we partition $\BraceTwo{N}$ into $N/16$ consecutive groups, each containing 16 numbers. By denoting them with $G_i$, where $i=0,1,\cdots, N/16-1$, we have $G_0 = \BraceTwo{16}$ and $G_i = \{16i + g : g\in G_0\}$. Further, observe that $Q_j$ of all the elements in $G_i$ have exactly the same value of $T_2$. As such, inclusion of $j \in G_i$ in $\tilde{A}_i$ depends solely on $T_1$, or the last four bits of $\langle j \rangle_2$. Letting $\widetilde{j}$ denote the decimal number corresponding to the last 4 bits of $j$, we observe that the values of $T_1$ decreases monotonically when $\widetilde{j}$ takes on values from $Q_0^{15} = (15,14,13,11,7,12,10,9,6,5,3,8,4,2,1,0)$ left to right. As such, for Case $k$ ($k=0,1,\cdots,16$), the first $k$ bit-channel indexes are selected in $Q_{0}^{15}$ for information transfer, and the rest are frozen to zero. Consequently, only 17 unique cases will occur, and the afore-mentioned four extra cases will not occur.
\end{IEEEproof}

\section{Results}
\label{sec:Results}

In this section, we compare the proposed decoding strategy with the SC decoder in terms of the bit-error-rate  and decoding latency performances.
\subsection{BER Performance}
\begin{figure}
  \centering
  \includegraphics[width=\columnwidth]{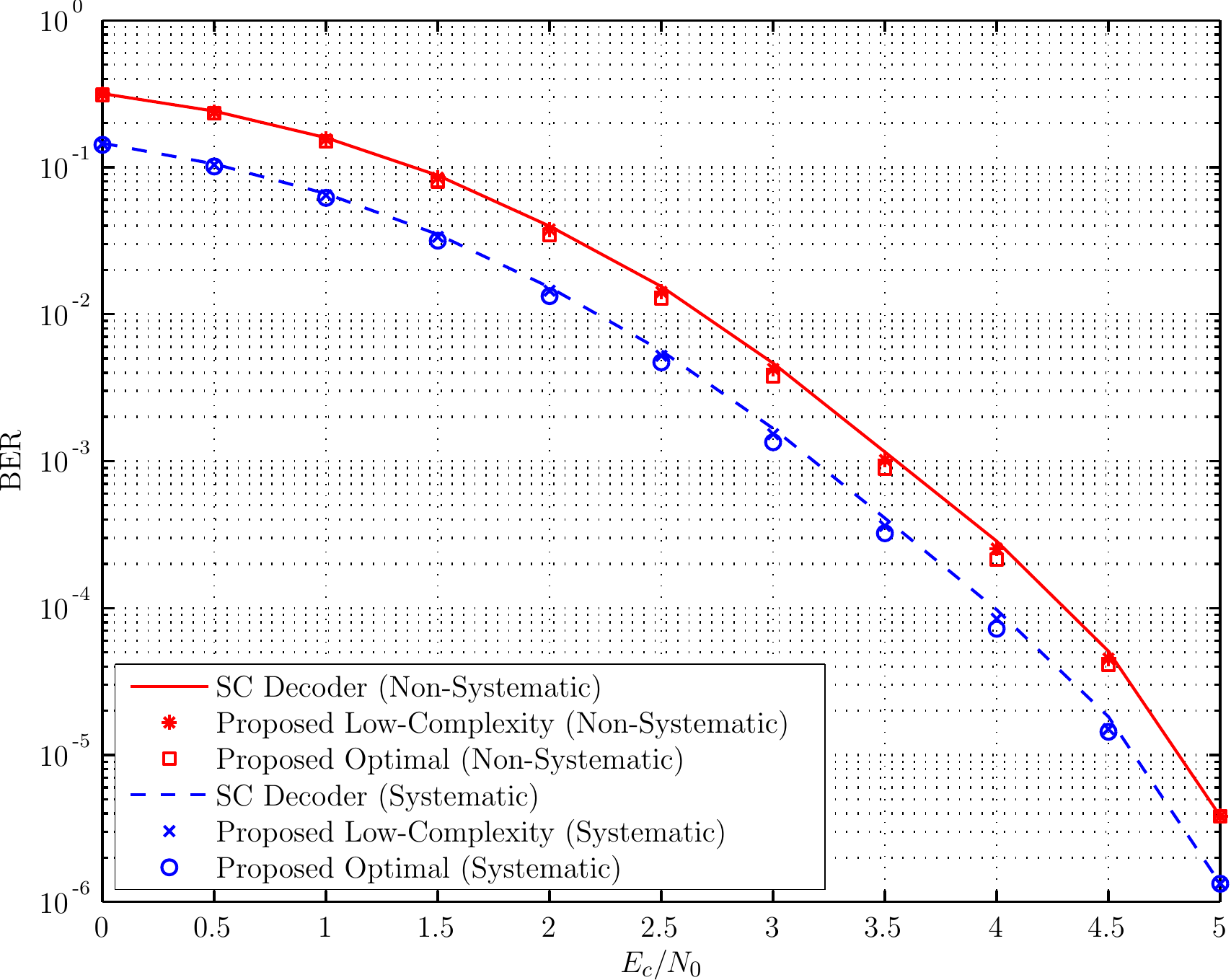}\\
  \caption{BER performance of systematic and non-systematic rate-1/2 polar codes of length $N=256$  in additive-white-Gaussian-noise channel with the signal-to-noise ratio of $E_c/N_0$ when decoded by conventional SC decoder and the proposed decoders for $R=8$. \label{Fig:BERR8SCOptLowCompl}}
\end{figure}

Fig. \ref{Fig:BERR8SCOptLowCompl} compares the BER performance of the proposed 8-bit parallel decoders with that of the SC decoder. Both optimal and low-complexity sub-optimal decoders are implemented for the proposed decoders. Here, the polar codes are constructed for the binary-erasure channel with the erasure probability of $e^{-1} \simeq 0.37$ \cite{PolarRMCompare,AWGNComparative}.

Some interesting observations can be made by analyzing the figure. First, the proposed decoders do not deteriorate the performance of the SC decoder. Second, the performance gap between the optimal and the sub-optimal decoders is negligibly small, which implies that the low-complexity decoders can be used instead of the optimal ones.
Last, the proposed schemes can be used for both systematic and non-systematic polar codes.

\subsection{Decoding Latency}

One way to approximate the decoding latency of different polar decoders is as follows. We presume that bit operations and addition/subtraction of real numbers can be carried out in one clock cycle, whereas a check-node operation takes $T_c$ and finding the minimum of a list takes $T_m$ clock cycles. Note that finding a minimum requires significantly less computations than performing a check-node operation \cite{LDPCRedComplexity}.

For a block of $R$ bits, the SC decoder performs $R/2$ check-node operations in each of $\log_2(R)$ stages. But all the check-node operations can be performed in parallel in the first stage, whereas the last stage requires all check-node operations to be performed in a sequential manner. In general, the number of parallel check-node operations performed in the $m$th stage is $R/2^m$, where $m=1, 2, \cdots, \log_2(R)$. Consequently, a decoding latency of $T_c(R/2 + R/4 + \cdots +1)=(R-1)T_c$ cycles is incurred in performing check-node operations for the SC decoder. Similarly, the number of binary and real additions involved in decoding $R$ bits be shown to be 1 and $R-1$, respectively. Therefore, an SC decoder will take $R+(R-1)T_c$ clock cycles to decode a block of $R$ bits. Note that this decoding latency is fixed regardless of the frozen-bit sequence.

For the proposed schemes, the decoding latency varies depending on the frozen-bit sequence. For example, for $R=8$, Case 0 can be executed instantaneously, whereas Case 4 incurs the highest decoding latency of $1+\max{\{T_c,T_m\}}$ cycles, which is calculated as follows. The check-node operations are performed is parallel, so they take only $T_c$ cycles for execution. Meanwhile, two outputs can be generated corresponding to $x_8=0$ and $1$ by two Wagner decoders ($T_m+1$ clock cycles), and, depending on the value of $x_8$ (computation time is 1 cycle), one of the outputs is selected. Therefore, the decoding latency for Case 4 is $\max{\{1+T_c,1+T_m\}}$.

Observe that even the highest decoding latency of the proposed decoders is much smaller than that of the SC decoder. Hence, our proposed scheme will significantly improve the decoding speed of variable-rate polar codes.

\begin{table*}[!htb]
\renewcommand{\arraystretch}{1.3}
\caption{Decoding latencies of the proposed low-complexity decoders for $R=8$ and $R=16$.}\label{DecodeLatency}
\centering
\begin{tabular}{|l|l|l|l|l|l|}
  \hline
  $R=8$ & Latency, $L$ & $R=16$ & Latency, $L$ & $R=16$ & Latency, $L$ \\ \hline
  Case 0 & 0 & Case 0 & 0 & Case 8 &  $1+\max{\{T_c,T_m\}}$ \\ \hline
  Case 1 & 1 & Case 1 & 1 & Case 9 &  $1+\max{\{T_c,T_m\}}$ \\ \hline
  Case 2 & 1 & Case 2 & 1 & Case 10 & $3+T_c+T_m$ \\ \hline
  Case 3 & 1+$T_m$ & Case 3 & 1+$T_m$ & Case 11 & $3+T_c+T_m+\max{\{T_c,T_m\}}$ \\ \hline
  Case 4 & $1+\max{\{T_c,T_m\}}$ & Case 4 & $2+\max{\{T_c,T_m\}}$ & Case 12 & $3+T_c+\max{\{T_c,T_m\}}$ \\ \hline
  Case 5 & $1+T_c$ & Case 5 & $3+\max{\{T_c,T_m\}}$ & Case 12 & $T_m$ \\ \hline
  Case 6 & $T_m$ & Case 6 & $1+T_m$ & Case 13 & $\max{\{1+T_c,T_m\}}$ \\ \hline
  Case 7 & $T_m$ & Case 6 & $3+T_c$ & Case 14 & $T_m$ \\ \hline
  Case 8 & 0 & Case 7 & $1+T_m$ & Case 15 & $T_m$ \\ \hline
  \multicolumn{2}{|l|}{}
             & Case 7 & $2+T_m$ & Case 16 & 0 \\ \cline{3-6}
  \multicolumn{2}{|l|}{}
             & Case 8 & $2+T_m$ &  \multicolumn{2}{l|}{} \\ \cline{3-6}

  \hline
\end{tabular}
\end{table*}

Table \ref{DecodeLatency} shows the decoding latency, $L$, of the proposed low-complexity decoders for blocks of length $R=8$ and $R=16$ bits. Observe that the proposed decoders have significantly less decoding latency compared to the SC decoder, which has the decoding latency of $R+(R-1)T_c$ cycles for all the cases.

\section{Conclusion}
\label{sec:Conclusion}
In this work, we presented fast 8-bit and 16-bit parallel decoders that can reduce the decoding-tree depth of the decoding tree of variable-rate and variable-length polar codes. They can reduce both the decoding latency and hardware complexity without deteriorating the bit-error-rate performance of the successive-cancellation decoder.

\appendix[Block Size 16 Decoders]
\label{Appendix:DecodersR16}
This appendix details the proposed low-complexity parallel decoders for the cases tabulated in Table \ref{AllCasesR16}.
\subsubsection{Case 0}
The optimal decoder assigns an all-zero vector to the output.
\subsubsection{Case 1}
This is a repetition code, and the optimal decoder makes a hard decision on the sum of the LLRs of the received bits.
\subsubsection{Case 2}
This is a (16,2) repetition code, and the optimal ML estimates of $\x{e}$ and $\x{f}$ can be found by making hard-decisions on the LLR sums of even-indexed and odd-indexed bits, respectively.
\subsubsection{Case 3}
This is a (4,3) SPC code concatenated with a (16,4) repetition code. The optimal decoder will first add the LLRs of the received bits corresponding to $\x{def}$, $\x{d}$, $\x{e}$, and $\x{f}$ before finding their hard estimates with the Wagner decoder.
\subsubsection{Case 4}
This is an (8,4) extended Hamming code concatenated with a (16,8) repetition code. Decoders mentioned in Case 4 for $R=8$ can be used after adding the LLRs of the first half to the second's.
\subsubsection{Case 5}
Although an exhaustive-search based ML decoder can be implemented, we can reduce the decoding complexity by introducing a new variable, $z=\x{7}+\x{f}$. By noting that $\x{0}+\x{8}=\x{1}+\x{9}=\cdots=\x{7}+\x{f}=z$, we first find $\widehat{z}$, a hard-decision estimate of $z$. Specifically, we add $y_0 \boxast y_8, y_1 \boxast y_9, \cdots y_7 \boxast y_{\textsc{f}}$ to get an LLR for $z$ and make a hard decision on the LLR to compute $\widehat{z}$.

Afterwards, the decoder computes $y_0 \pm y_{8},y_{1} \pm y_{9}, \cdots,y_7 \pm y_{\textsc{f}}$, where addition is performed when $\widehat{z}=0$, and subtraction otherwise. These values are then input to one of the decoders of Case 4 for $R=8$ to get estimates of $\x{8}, \x{9}, \cdots, \x{f}$. Lastly $\widehat{z}$ is added to $\widehat{\x{8}}, \widehat{\x{9}}, \cdots, \widehat{\x{f}}$ to compute $\widehat{\x{0}}, \widehat{\x{1}}, \cdots, \widehat{\x{7}}$.

Implementing two parallel decoders corresponding to $\widehat{z}$ equalling 0 and 1 and choosing an appropriate output after computing $\widehat{z}$ can reduce the decoding latency.
\subsubsection{Case 6}
For the first case, $\mathbf{f}=\mathrm{FFC0}$, an optimal decoder can be implemented by noting that the codeword comprises two concatenated (4,3) SPC and (8,4) repetition codes. Two separate Wagner decoders can be used to find hard estimates of the transmitted bits after adding the LLRs of the repeated bits.

For the second case, $\mathbf{f}=\mathrm{FEE0}$, the decoder presented in Case 5 ($R=16$) can be used to decode the received LLRs with the exception that the decoder of Case 4 ($R=8$) is replaced with the low-complexity decoder of Case 5 ($R=8$).

\subsubsection{Case 7}
The first case, $\mathbf{f}=\mathrm{FF80}$, corresponds to a concatenated (8,7) SPC and (16,8) repetition code. Therefore, the code can be decoded optimally by adding the LLRs of $\x{0},\x{1},\cdots,\x{7}$ to that of $\x{8},\x{9},\cdots,\x{f}$ and finding hard estimates by the Wagner decoder.

For the second case, $\mathbf{f}=\mathrm{FEC0}$, the decoder presented in Case 5 ($R=16$) can be used to decode the received LLRs with the exception that the low-complexity decoder of Case 6 ($R=8$) replaces the decoder of Case 4 ($R=8$).

\subsubsection{Case 8}
For the first case, $\mathbf{f} = \mathrm{FE80}$, the decoder presented in Case 5 ($R=16$) can be used to decode the received LLRs with the exception that the Wagner decoder is used instead of the decoder of Case 4 ($R=8$).

For the second case, $\mathbf{f}=\mathrm{FCC0}$, the even-indexed and the odd-indexed bits constitute two separate (8,4) extended Hamming codes. Therefore, the decoders for Case 4 ($R=8$) can be used to decode the received code vector.
\subsubsection{Case 9}
A low-complexity decoder can be implemented by defining $z_0=\x{6}+\x{e}$ and $z_1=\x{7}+\x{f}$. Observe that $\x{0}+\x{8}=\x{2}+\x{a}=\x{4}+\x{c}=\x{6}+\x{e}=z_0$, and $\x{1}+\x{9}=\x{3}+\x{b}=\x{5}+\x{d}=\x{7}+\x{f}=z_1$. The proposed decoder makes hard decisions on $y_0 \boxast y_8 + \cdots + y_6 \boxast y_{\textsc{e}}$ and $y_1 \boxast y_9 + \cdots + y_7 \boxast y_{\textsc{f}}$ and respectively assigns them to $\widehat{z}_0$ and $\widehat{z}_1$, hard estimates of $z_0$ and $z_1$, respectively. Afterwards, additional LLRs for $\x{8}, \x{9}, \cdots, \x{f}$ are computed by using $\widehat{z}_0$ and $\widehat{z}_1$. For example, the additional LLR of $\x{8}$ is $y_0$ when $\widehat{z}_0$ is 0 and $-y_0$ otherwise. After adding the additional LLRs to $y_{8}, y_{9}, \cdots, y_{\textsc{f}}$, Wagner decoder is used to find hard estimates of $\x{8}, \x{9}, \cdots, \x{f}$, which along with $\widehat{z}_0$ and $\widehat{z}_1$ are used to estimate $\x{0}, \x{1}, \cdots, \x{7}$.

The decoding latency can be reduced by implementing four Wagner decoders (corresponding to four possible values of $(\widehat{z}_0,\widehat{z}_1)$) and using the computed value of $(\widehat{z}_0,\widehat{z}_1)$ to select the output of corresponding Wagner decoder.

\subsubsection{Case 10}
A low-complexity decoder can be implemented by defining four variables: $z_0=\x{0}+\x{8}=\x{4}+\x{b}$, $z_1=\x{1}+\x{9}=\x{5}+\x{c}$, $z_2=\x{2}+\x{a}=\x{6}+\x{d}$, and $z_3=\x{3}+\x{b}=\x{7}+\x{f}$. The decoder will first compute $\widehat{z}_i$'s, hard estimates of $z_i$'s, where $i=0,1,2,3$, from the LLRs obtained by adding LLRs of the output of check-node operations. For example, LLR of $z_0$ is computed by adding $y_{0} \boxast y_{8}$ and $y_{4} \boxast y_{\textsc{b}}$, where $y_i$ denotes the LLR of $x_i$. Depending on the value of $z_i$'s, additional LLRs for $\x{8}, \x{9}, \cdots, \x{f}$ are obtained from $y_0, y_1, \cdots, y_7$. For example, additional LLR for $\x{8}$ is $y_0$ when $z_0=0$ and is $-y_0$ when $z_0=1$. After adding the additional LLRs to the received LLRs, the decoder finds hard estimates of $\x{8}, \x{9}, \cdots, \x{f}$ using Wagner decoder. Finally, these hard estimates are used along with $\widehat{z}_i$'s to estimate $\x{0}, \x{1}, \cdots, \x{7}$.

\subsubsection{Case 11}
A low-complexity decoder can be implemented by observing that $z_0, z_1, \cdots, z_7$ constitute an (8,4) extended Hamming code, where $z_0 = \x{0}+\x{8} , z_1 = \x{1}+\x{9},\cdots, z_7 = \x{7}+\x{f}$. As such, the decoders of Case 4 for $R=8$ can be used to find $\widehat{z}_i$'s, estimates of $z_i$'s for $i=0, 1, \cdots, 7$. Then, depending on the values of $\widehat{z}_i$, additional LLRs for $\x{8}, \x{9}, \cdots, \x{f}$ are obtained from $y_i$'s, where $i=0,1, \cdots, 7$. For example, the additional LLR for $\x{8}$ is $y_0$ if $\widehat{z}_0=0$ and $-y_0$ otherwise. After adding the LLRs, Wagner decoder is used to compute estimates of $\x{8}, \x{9}, \cdots, \x{f}$. The decoded bits along with $\widehat{z}_0, \widehat{z}_1, \cdots, \widehat{z}_7$ are then used to compute estimates of $\x{0},\x{1},\cdots,\x{7}$.

\subsubsection{Case 12}
For the first case, $\mathbf{f}=\mathrm{E800}$, the decoder of Case 11 can be used to decode the received LLRs, with the exception that the Wagner decoder is not used at all. Rather, hard decisions are made on the updated LLRs of  $\x{8}, \x{9}, \cdots, \x{f}$. 

In the second case, $\mathbf{f}=\mathrm{C0C0}$, the code word consists of four separate (4,3) SPC codes, which can be individually decoded by the Wagner rule.

\subsubsection{Case 13}
\begin{figure}
  \centering
  \includegraphics[width=0.5\columnwidth]{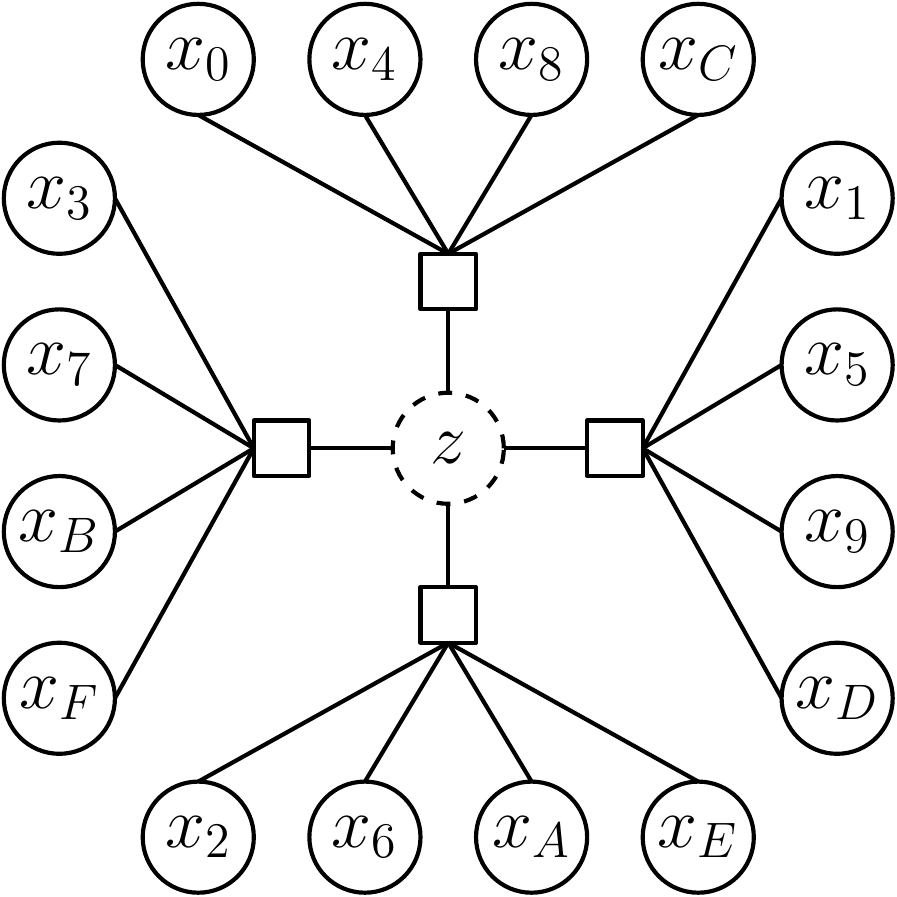}\\
  \caption{Tanner graph for the systematic-polar code corresponding to $\mathbf{f}=\mathrm{E000}$.}\label{Case13TannerGraph}
\end{figure}

Introducing a new variable, $z = \x{3}+\x{7}+\x{b}+\x{f}$, results in a cycle-free Tanner graph as shown in Fig. \ref{Case13TannerGraph}. As such, a non-iterative optimal MAP decoder can easily be implemented for this case.

A low-complexity decoder can also be constructed for this code. For example, making a hard decision on $z$ results in four separate SPC codes, which can be decoded by the Wagner rule.

\subsubsection{Case 14}
This code consists of two (8,7) SPC codes, which can be optimally decoded by two Wagner decoders.
\subsubsection{Case 15}
This is a (16,15) SPC code, and Wagner decoder can be used to decode the received LLRs optimally.
\subsubsection{Case 16}
The optimal decoder will make hard decisions on the LLRs of the received bits.
\balance


\begin{thebibliography}{10}
\providecommand{\url}[1]{#1}
\csname url@samestyle\endcsname
\providecommand{\newblock}{\relax}
\providecommand{\bibinfo}[2]{#2}
\providecommand{\BIBentrySTDinterwordspacing}{\spaceskip=0pt\relax}
\providecommand{\BIBentryALTinterwordstretchfactor}{4}
\providecommand{\BIBentryALTinterwordspacing}{\spaceskip=\fontdimen2\font plus
\BIBentryALTinterwordstretchfactor\fontdimen3\font minus
  \fontdimen4\font\relax}
\providecommand{\BIBforeignlanguage}[2]{{%
\expandafter\ifx\csname l@#1\endcsname\relax
\typeout{** WARNING: IEEEtran.bst: No hyphenation pattern has been}%
\typeout{** loaded for the language `#1'. Using the pattern for}%
\typeout{** the default language instead.}%
\else
\language=\csname l@#1\endcsname
\fi
#2}}
\providecommand{\BIBdecl}{\relax}
\BIBdecl

\bibitem{PolarCodesSeminalWork}
E.~Ar{\i}kan, ``Channel polarization: A method for constructing
  capacity-achieving codes for symmetric binary-input memoryless channels,''
  \emph{{IEEE} Trans. Inf. Theory}, vol.~55, no.~7, pp. 3051--3073, Jul. 2009.

\bibitem{FastPolarDecoders}
G.~Sarkis, P.~Giard, A.~Vardy, C.~Thibeault, and W.~J. Gross, ``Fast polar
  decoders: Algorithm and implementation,'' \emph{{IEEE} J. Sel. Areas
  Commun.}, vol.~32, no.~5, pp. 946--957, May 2014.

\bibitem{SarkisSystematicPolarCodes}
G.~Sarkis, I.~Tal, P.~Giard, A.~Vardy, C.~Thibeault, and W.~J. Gross,
  ``Flexible and low-complexity encoding and decoding of systematic polar
  codes,'' \emph{{IEEE} Trans. Commun.}, vol.~64, no.~7, pp. 2732--2745, Jul.
  2016.

\bibitem{UnrolledPolarDec}
P.~Giard, G.~Sarkis, C.~Thibeault, and W.~J. Gross, ``237 {Gbit/s} unrolled
  hardware polar decoder,'' \emph{Electron. Lett.}, vol.~51, no.~10, pp.
  762--763, May 2015.

\bibitem{SimplifiedSCD}
A.~Alamdar-Yazdi and F.~R. Kschischang, ``A simplified successive-cancellation
  decoder for polar codes,'' \emph{{IEEE} Commun. Lett.}, vol.~15, no.~12, pp.
  1378--1380, Dec. 2011.

\bibitem{PolarCodesMagazine}
K.~Niu, K.~Chen, J.~Lin, and Q.~T. Zhang, ``Polar codes: Primary concepts and
  practical decoding algorithms,'' \emph{{IEEE} Commun. Mag.}, vol.~52, no.~7,
  pp. 192--203, Jul. 2014.

\bibitem{HuaweiPolarCodes}
Huawei and HiSilicon, ``Details of the polar code design,'' 3GPP TSG RAN WG1
  Meeting\#87, Reno, USA, Tech. Rep. R1-1611254, Nov. 2016.

\bibitem{QualcommLDPCCodes}
{Qualcomm Incorporated}, ``{LDPC} rate and compatible design overview,'' 3GPP
  TSG-RAN WG1 \#86bis, Lisbon, Portugal, Tech. Rep. R1-1610137, Oct. 2016.

\bibitem{EricssonPolarCode}
Ericsson, ``Performance study of polar code candidates,'' 3GPP TSG-RAN WG1
  \#88, Athens, Greece, Tech. Rep. R1-1703538, Feb. 2017.

\bibitem{TwoBitParallelDecoder}
B.~Yuan and K.~K. Parhi, ``Low-latency successive-cancellation polar decoder
  architectures using 2-bit decoding,'' \emph{{IEEE} Trans. Circuits Syst.
  {I}}, vol.~61, no.~4, pp. 1241--1254, Apr. 2014.

\bibitem{MultiBitParallelDecWithoutLLR}
------, ``Low-latency successive-cancellation list decoders for polar codes
  with multibit decision,'' \emph{{IEEE} Trans. {VLSI} Syst.}, vol.~23, no.~10,
  pp. 2268--2280, Oct. 2015.

\bibitem{MultiBitParallelDecoder}
------, ``{LLR}-based successive-cancellation list decoder for polar codes with
  multibit decision,'' \emph{{IEEE} Trans. Circuits Syst. {II}}, vol.~64,
  no.~1, pp. 21--25, Jan. 2017.

\bibitem{PartiallyParallelPolarEncoder}
H.~Yoo and I.~C. Park, ``Partially parallel encoder architecture for long polar
  codes,'' \emph{{IEEE} Trans. Circuits Syst. {II}}, vol.~62, no.~3, pp.
  306--310, Mar. 2015.

\bibitem{ArikanSystPolarCodes}
E.~Ar{\i}kan, ``Systematic polar coding,'' \emph{{IEEE} Commun. Lett.},
  vol.~15, no.~8, pp. 860--862, Aug. 2011.

\bibitem{DualPolarCode}
N.~Hussami, S.~B. Korada, and R.~Urbanke, ``Performance of polar codes for
  channel and source coding,'' in \emph{IEEE Int. Symp. Inf. Theory}, Jun.
  2009, pp. 1488--1492.

\bibitem{PermutedSCD}
H.~Vangala, E.~Viterbo, and Y.~Hong, ``Permuted successive cancellation decoder
  for polar codes,'' in \emph{Int. Symp. Inf. Theory Applicat.}, Oct. 2014, pp.
  438--442.

\bibitem{PolarCodePermutationThesis}
\BIBentryALTinterwordspacing
J.~Guo, ``Polar codes for reliable transmission: Theoretical analysis and
  applications,'' Ph.D. dissertation, University of Cambridge, Jun. 2015.
  [Online]. Available:
  \url{http://itc.upf.edu/system/files/biblio-pdf/Thesis_JingGuo.pdf}
\BIBentrySTDinterwordspacing

\bibitem{MLSoftDecodBlockCode}
J.~Snyders and Y.~Be'ery, ``Maximum likelihood soft decoding of binary block
  codes and decoders for the {Golay} codes,'' \emph{{IEEE} Trans. Inf. Theory},
  vol.~35, no.~5, pp. 963--975, Sep. 1989.

\bibitem{WagnerDecoder}
R.~Silverman and M.~Balser, ``Coding for constant-data-rate systems,''
  \emph{IRE Trans. Prof. Group Inform. Theory}, vol.~4, no.~4, pp. 50--63, Sep.
  1954.

\bibitem{ErrorControlDigCommWicker}
S.~B. Wicker, \emph{Error control systems for digital communication and
  storage}.\hskip 1em plus 0.5em minus 0.4em\relax Prentice hall Englewood
  Cliffs, 1995, vol.~1.

\bibitem{CycleFreeFactorGraph}
F.~R. Kschischang, B.~J. Frey, and H.~A. Loeliger, ``Factor graphs and the
  sum-product algorithm,'' \emph{{IEEE} Trans. Inf. Theory}, vol.~47, no.~2,
  pp. 498--519, Feb. 2001.

\bibitem{PolarRMCompare}
E.~Ar{\i}kan, ``A performance comparison of polar codes and {Reed-Muller}
  codes,'' \emph{{IEEE} Commun. Lett.}, vol.~12, no.~6, pp. 447--449, Jun.
  2008.

\bibitem{AWGNComparative}
\BIBentryALTinterwordspacing
H.~Vangala, E.~Viterbo, and Y.~Hong, ``A comparative study of polar code
  constructions for the {AWGN} channel,'' \emph{arXiv preprint
  arXiv:1501.02473}, 2015. [Online]. Available:
  \url{http://arxiv.org/abs/1501.02473}
\BIBentrySTDinterwordspacing

\bibitem{LDPCRedComplexity}
M.~P.~C. Fossorier, M.~Mihaljevic, and H.~Imai, ``Reduced complexity iterative
  decoding of low-density parity check codes based on belief propagation,''
  \emph{{IEEE} Trans. Commun.}, vol.~47, no.~5, pp. 673--680, May 1999.

\end{thebibliography}
\end{document}